# Optimization-Based Image Reconstruction Regularized with Inter-Spectral Structural Similarity for Limited-Angle Dual-Energy Cone-Beam CT

Junbo Peng[1], Tonghe Wang[2], Huiqiao Xie[2], Richard L. J. Qiu[1], Chih-Wei Chang[1], Justin Roper[1], David S. Yu[1], Xiangyang Tang[3] and Xiaofeng Yang[1*]

[1]Department of Radiation Oncology and Winship Cancer Institute, Emory University, Atlanta, GA 30322
[2]Department of Medical Physics, Memorial Sloan Kettering Cancer Center, New York, NY 10065
[3]Department of Radiology and Imaging Sciences and Winship Cancer Institute, Emory University, Atlanta, GA 30322
*Email: xiaofeng.yang@emory.edu

**Running title**: Dual-Energy CBCT Imaging

**Manuscript Type:** Original Research

## Abstract

**Purpose:** Limited-angle dual-energy (DE) cone-beam CT (CBCT) is considered as a potential solution to achieve fast and low-dose DE imaging on current CBCT scanners without hardware modification. However, its clinical implementations are hindered by the challenging image reconstruction from limited-angle projections. While optimization-based and deep learning-based methods have been proposed for image reconstruction, their utilization is limited by the requirement for X-ray spectra measurement or paired datasets for model training. This work aims to facilitate the clinical applications of fast and low-dose DE-CBCT by developing a practical solution for image reconstruction in limited-angle DE-CBCT.

**Methods:** An inter-spectral structural similarity-based regularization was integrated into the iterative image reconstruction in limited-angle DE-CBCT. By enforcing the similarity between the DE images, limited-angle artifacts were efficiently reduced in the reconstructed DECBCT images. The proposed method was evaluated using two physical phantoms and three digital phantoms, demonstrating its efficacy in quantitative DECBCT imaging.

**Results:** In all the studies, the proposed method achieves accurate image reconstruction without visible residual artifacts from limited-angle DE-CBCT projection data. In the digital phantom studies, the proposed method reduces the mean-absolute-error (MAE) from 309/290 HU to 14/20 HU, increases the peak signal-to-noise ratio (PSNR) from 40/39 dB to 70/67 dB, and improves the structural similarity index measurement (SSIM) from 0.74/0.72 to 1.00/1.00.

**Conclusions:** The proposed method achieves accurate optimization-based image reconstruction in limited-angle DE-CBCT, showing great practical value in clinical implementations of limited-angle DE-CBCT.

## 1. Introduction

Dual-energy (DE) CT, as an implementation of spectral CT [1], has found a variety of clinical applications, including virtual non-contrast imaging [2-4], virtual monochromatic imaging [5], automatic bone removal [6], and iodine quantification [7]. On the other hand, cone-beam CT (CBCT), using a flat-panel detector (FPD), is a volumetric imaging technique that plays a vital role in image-guided radiation therapy [8], surgery [9], and interventional procedures [10]. Recently, integrating DE imaging with CBCT has attracted increasing attention. It has shown promise in various applications, such as contrast-enhanced lesion detection [11], angiography [12], material classification [13], and dose calculation for image-guided photon and proton therapy [14-16].

However, the routine practice of DE-CBCT has been hindered due to the lack of a practical DE-CBCT implementation. Early efforts in DE-CBCT involved performing two separate scans at different tube voltages [13]. Besides the excessive radiation dose compared to a single-energy scan, the utility of this data acquisition scheme is hindered by the motion-induced spatial misregistration between the two scans because of the much slower CBCT gantry rotation than that of diagnostic CT due to patient safety considerations[17]. To achieve DE-CBCT with data acquisition time and radiation dose comparable to the single-energy scans, techniques such as fast kVp-switching and dual-layer FPD have been investigated on CBCT systems [12, 18-20]. Nonetheless, the fast kVp-switching faces technical challenges due to instability in the rapid X-ray tube voltage modulation on current CBCT scanners [21], and the dual-layer FPD suffers from limited DE spectral separation due to its physical structure [19]. In short, despite the success of these two schemes in diagnostic DECT scanners, both of them require sophisticated hardware upgrades on existing onboard CBCT scanners and have not been commercialized yet.

It has been an active research area to achieve low-dose single-scan DE imaging on current scanners without significant hardware modifications since the introduction of the slow kVp-switching technique with prior image constrained compressed sensing (PICCS)-based image reconstruction [22]. From the data acquisition perspective, the reported methods can be divided into spatial and temporal downsampling schemes. The spatial downsampling strategies install a reciprocating or static beam filter on the source or detector side to selectively alter the spectrum of X-ray photons in specific paths, [17, 21, 23-26]. generating spatially alternated high- and low-energy projection data in each projection view. Two common challenges of these methods are partial filtration in the penumbra region and increased noise after the beam filtration. The temporal downsampling methods acquire DE projection data with alternating angular distribution via a slow kVp-switching scheme or a fast-rotating filter installed on the source side [22, 27-30]. Among all the proposed data acquisition schemes, limited-angle DE-CBCT is the most cost-effective and feasible-to-implement solution in terms of hardware, and it is readily available for existing scanners [28, 29, 31]. As a temporal downsampling strategy, limited-angle DE-CBCT requires no hardware modification and alters the tube voltage only once in the middle of the source trajectory, acquiring DE projection data within two complementary limited-angle arcs. Besides, the spectral separation between the DE projection data in this scheme can be higher than that of the beam filter-based solutions. In brief, limited-angle DE-CBCT is an optimal candidate for practical DE-CBCT solutions on existing scanners.

One of the major challenges of limited-angle DE-CBCT is reconstruction artifacts. Optimization-based reconstruction algorithms and deep learning models have been proposed to address this issue. The optimization-based methods exploit constraints on the image directional-total-variation (DTV) of virtual monochromatic or material-specific images during the material decomposition [31-34]. Such iterative material decomposition algorithms require accurate measurement or estimation of DE X-ray spectra, which is challenging on current CBCT scanners. Furthermore, the optimization in these algorithms is computationally intensive due to the integration of material decomposition. Ignoring imaging physics, deep learning models directly learn the mapping from the acquired data to the desired data via supervised learning. The data restoration can be performed in the projection domain or the image domain. The projection-domain model translates the under-sampled DE projection data to the full-sampled data, which are used for subsequent analytical reconstruction [29]. The image-domain model translates the directly reconstructed DE-CBCT images contaminated by limited-angle artifacts to high-quality images without artifacts [28]. Despite the state-of-the-art performance of the deep learning-based methods, their clinical potential is significantly limited by the requirement of paired limited-angle and full-sampled DE-CBCT datasets for the supervised model training, which are unavailable in real patient datasets. In short, the clinical application of limited-angle DE-CBCT is hindered by the lack of a practical solution to the challenging image reconstruction.

In this study, we propose an efficient optimization-based image reconstruction method for limited-angle DE-CBCT to facilitate the clinical application of quantitative DE-CBCT. Recognizing the fact that the complete structural information is preserved in the projection data and the DE images share the same anatomical structures, we introduce an inter-spectral structural similarity-based regularization term into the iterative image reconstruction in limited-angle DE-CBCT. We demonstrate the feasibility of the proposed image reconstruction algorithm and evaluate its performance via a study of four physical phantoms and a study of three digital phantoms.

## 2. Materials and methods

### 2.1 Principles of iterative CT reconstruction

Iterative image reconstruction methods formulate the CT projection as a discretized linear model [35]

$$\vec{b} = F\vec{\mu} \tag{1}$$

where $\vec{\mu}$ denotes the vectorized CT image, $F$ is the system matrix representing the forward projection operator, and $\vec{b}$ indicates the acquired projection data. $\vec{\mu}$ has a dimension of $N_{vox} \times 1$ where $N_{vox}$ is the number of voxels in the CT image, $F$ is in a size of $N_{vox} \times N_v N_d$ where $N_v$ is the number of projection views and $N_d$ is the number of detector pixels, and $\vec{b}$ has a dimension of $N_v N_d \times 1$.

With a Gaussian noise assumption on the projection data, the maximum likelihood estimation (MLE) is equivalent to a least-square optimization problem

$$\min_{\vec{\mu}} \left\{ \frac{1}{N_v N_d} \left\| F\vec{\mu} - \vec{b} \right\|_2^2 + \mathcal{R}(\vec{\mu}) \right\} \quad s.t.\ \vec{\mu} \geqslant \vec{0} \tag{2}$$

where $\|\cdot\|_2^2$ is the $\ell_2$ norm measuring the Eculidean distance, and $\mathcal{R}(\cdot)$ is an image prior-based regularization term. For example, the total variation (TV) norm, $\|\cdot\|_{TV}$, is a common choice of $\mathcal{R}(\cdot)$ based on the piece-wise constant property of CT images [36].

### 2.2 Inter-spectral structural similarity-based iterative image reconstruction for limited-angle DE-CBCT

In limited-angle DE-CBCT, the high-kVp projection data are acquired within the first half of the view angular range, and the low-kVp projection data is acquired within the other half. Thus, the data acquisition in limited-angle DE-CBCT can be formulated as

$$\begin{cases} \vec{b}_H = F_H \vec{\mu}_H \\ \vec{b}_L = F_L \vec{\mu}_L \end{cases} \tag{3}$$

where $\vec{\mu}_{H,L}$ are the DE-CBCT images, $F_{H,L}$ represent the forward projection matrices in the high- and low-kVp scans, and $\vec{b}_{H,L}$ indicate the acquired high- and low-kVp projection data.

$F_H$ and $F_L$ are the top and bottom half of the system matrix in a conventional single-energy scan in Eq. (1), i.e.,

$$F = \begin{bmatrix} F_H \\ --- \\ F_L \end{bmatrix} \tag{4}$$

The projection data $\vec{b}_{H,L}$ at each X-ray spectrum are angular-limited, leading to severe limited-angle artifacts in the DE-CBCT images directly reconstructed from the acquired projection data.

The proposed iterative method for limited-angle DE-CBCT reconstruction is based on the fact that the complete anatomical information acquired in the mixed-spectra projection data and the anatomical structures are consistent between DE-CBCT images. Therefore, we introduce a structural similarity-based regularization term to suppress the limited-angle artifacts during the iterative reconstruction:

$$\min_{\vec{\mu}_H,\vec{\mu}_L} \left\{ \frac{1}{N_v N_d} \left[ \left\| F_H \vec{\mu}_H - \vec{b}_H \right\|_2^2 + \left\| F_L \vec{\mu}_L - \vec{b}_L \right\|_2^2 \right] + \lambda [1 - \mathrm{SSIM}(\vec{\mu}_H, \vec{\mu}_L)] \right\} \\ s.t.\ \vec{\mu}_{H,L} \succcurlyeq \vec{0} \tag{5}$$

where $\lambda$ is the user-defined weighting factor that balances the tradeoff between reconstruction data fidelity and image regularization. In Eq. (5), SSIM(·,·) calculates the structural similarity index measurement (SSIM) between two input images by [37]

$$\mathrm{SSIM}(\vec{\mu}_H, \vec{\mu}_L) = \frac{\left(2\overline{\vec{\mu}_H \vec{\mu}_L} + C_1\right)\left(2\sigma_{H,L} + C_2\right)}{\left(\overline{\vec{\mu}_H}^2 + \overline{\vec{\mu}_L}^2 + C_1\right)\left(\sigma_H^2 + \sigma_L^2 + C_2\right)} \tag{6}$$

where $\overline{\vec{\mu}_H}$ and $\overline{\vec{\mu}_L}$ are the mean values of DE-CBCT, $\sigma_H$ and $\sigma_L$ are the standard deviations of DE-CBCT, $\sigma_{H,L}$ indicates the covariance between DE-CBCT, and $C_{1,2}$ are two constants associated with the dynamic range of DE-CBCT.

The incorporation of the SSIM regularization is the key to the success of limited-angle artifact reduction in iterative DE-CBCT reconstruction, which is the major contribution of this work. Of note, conventional TV or DTV regularization can also be involved in the proposed iterative reconstruction for noise suppression.

### 2.3 Evaluation

The performance of the proposed iterative method for image reconstruction in limited-angle DE-CBCT has been evaluated using data from both physical and digital phantoms. The physical phantom data were acquired on the table-top CBCT system, and the digital phantom data were simulated using the DECT images acquired on a Siemens TwinBeam scanner.

#### 2.3.1 Physical phantom study

Due to the difficulty of raw data acquisition in the commercial scanners, we collected the DE projection data of two physical phantoms from the table-top CBCT system.

A Catphan©700 phantom and an anthropomorphic head phantom were scanned on one system. The geometry of the system was designed to match that of a Varian On-Board Imager (OBI) CBCT scanner on the TrueBeam radiation therapy machine. The source-to-detector distance (SDD) and the source-to-axis distance (SAD) are 1500 mm and 1000 mm, respectively. The FPD consists of 1408×1408 pixels with a size of 0.308×0.308 $mm^2$. A collimator was installed on the X-ray tube to suppress photon scattering. Therefore, only the central row of detector pixels was used for image reconstruction. In the low-energy data acquisition, the tube voltage was set to 125 kVp. In the high-energy data acquisition, the tube voltage was unchanged, but a 0.2-mm tin (Sn) filter was installed on the exit window of the X-ray tube, leading to a higher mean photon energy after the filtration. For both phantom scans, a total of 600 views were acquired over 360° angular range in these two scans. The limited-angle DE-CBCT data consists of 175 projection views out of 0° to 105° in the high-energy data and 175 views out of 105° to 210° in the low-energy data. The total angular range of 210° is based on the short scan protocol of a Varian OBI system [38]. The imaging parameters of these two phantom scans are summarized in Table I.

**Table I**. Imaging parameters in the limited-angle DE-CBCT scans of physical and digital phantoms.

| | **Physical Phantom Study** | | **Digital Phantom Study** |
|---|---|---|---|
| | **Catphan©700** | **Head Phantom** | |
| **High-Energy Spectrum** | 125 kVp + 0.2-mm Sn | | 120 kVp + 0.6-mm Sn |
| **Low-Energy Spectrum** | 125 kVp | | 120 kVp + 0.05-mm Au |
| **Reconstruction Volume** | 512×512×5 | | 256×256×150 |
| **Voxel Size ($mm^3$)** | 0.5×0.5×0.5 | | 1.0×1.0×1.0 |
| **SDD/SAD (mm)** | 1500/1000 | | |
| **Detector Dimension** | 1408×15 | | 1024×768 |
| **Detector Pixel Size ($mm^2$)** | 0.308×0.308 | | 0.388×0.388 |
| **High-Energy Projection Views** | 175 views from 0° to 105° | | |
| **Low-Energy Projection Views** | 175 views from 105° to 210° | | |

Of note, in the scans of physical phantoms, the DE spectral separation was achieved via beam filtration instead of different tube voltages. This is because the collected experimental data were originally acquired for other studies. The spectral separation in the digital phantom study is similar because of the hardware design of the Siemens TwinBeam scanner. In this work, we re-sorted and reused these data, which effectively preserved our research purpose.

2.3.2 Digital phantom study

The proposed method was further evaluated on three digital anthropomorphic phantoms based on DECT images of three H&N patients acquired on a Siemens TwinBeam DECT scanner at our institute. The tube voltage of the scanner was fixed at 120 kVp. The high- and low-energy data were acquired using a 0.6-mm tin filter and a 0.05-mm gold (Au) filter, respectively. For each patient, a volume of 256×256×150 with a voxel size of 1×1×1 $mm^3$ was selected for the simulated scan of limited-angle DE-CBCT.

First, the DECT images were converted from HU to linear attenuation coefficient (LAC) maps with the effective LACs of water at the high- and low-energy X-ray spectra, which were simulated using the Spektr toolbox [39]. Second, the limited-angle DE-CBCT scan was simulated based on the geometry of the OBI system in the physical phantom study, as listed in Table I. SDD and SID were 1500 and 1000 mm, and the FPD was 1024×768 with a pixel size of 0.388×0.388 $mm^2$. The first 175 projections within 0° to 105° were simulated using the high-energy CBCT volume, and the second 175 projections within 105° to 210° were simulated using the low-energy CBCT volume. The simulated forward projection was implemented using Siddon's ray-tracing algorithm [40].

2.3.3 Image quality metrics

In the physical phantom studies, the line profiles and HU measurements within regions-of-interest (ROIs) were used to evaluate the accuracy of DE-CBCT reconstruction. In addition, peak signal-to-noise ratio (PSNR) and SSIM were calculated to evaluate the distortion between the reconstructed images and the ground truth.

In the digital phantom studies, the accuracy of image reconstruction was evaluated using the mean absolute error (MAE) calculated within the body mask between the reconstructed and reference DE-CBCT images. The coherence between reconstructed and ground truth images was evaluated using PSNR and SSIM.

**2.4 Comparison studies**

The proposed method was compared with three other methods. The first directly performed Feldkamp-Davis-Kress (FDK) reconstruction using the limited-angle DE-CBCT projection data. The second one was DTV-regularized iterative reconstruction

$$\begin{gathered}\min_{\vec{\mu}_H}\left\{\frac{1}{N_{v,H}N_d}\left\|F_H\vec{\mu}_H-\vec{b}_H\right\|_2^2+\eta_i\|\vec{\mu}_H\|_{TV,i}+\eta_j\|\vec{\mu}_H\|_{TV,j}\right\}\\ \min_{\vec{\mu}_L}\left\{\frac{1}{N_{v,L}N_d}\left\|F_L\vec{\mu}_L-\vec{b}_L\right\|_2^2+\eta_i\|\vec{\mu}_L\|_{TV,i}+\eta_j\|\vec{\mu}_L\|_{TV,j}\right\}\\ s.t.\ \vec{\mu}_{H,L}\geqslant\vec{0}\end{gathered} \tag{7}$$

where the subscripts $i, j$ denote the directions for TV regularization, and $\eta_{i/j}$ indicates the weighting factor for each DTV penalty. Of note, the implementation of DTV-regularized image reconstruction here was different from that in the series of works[31-34], where the authors exploited the correlation between DE projection data via the material decomposition framework. In this work, the reconstructions of high- and low-energy images were independent, and only single-energy data was used for the reconstruction of each image.

The third comparison method was PICCS-based iterative reconstruction [22]

$$\begin{gathered}\min_{\vec{\mu}_H}\left\{\frac{1}{N_{v,H}N_d}\left\|F_H\vec{\mu}_H-\vec{b}_H\right\|_2^2+\eta\|\vec{\mu}_H\|_{TV}+\delta\|\vec{\mu}_H-\vec{\mu}_P\|_{TV}\right\}\\ \min_{\vec{\mu}_L}\left\{\frac{1}{N_{v,L}N_d}\left\|F_L\vec{\mu}_L-\vec{b}_L\right\|_2^2+\eta\|\vec{\mu}_L\|_{TV}+\delta\|\vec{\mu}_L-\vec{\mu}_P\|_{TV}\right\}\\ s.t.\ \vec{\mu}_{H,L}\geqslant\vec{0}\end{gathered} \tag{8}$$

where $\mu_P$ denotes the prior image reconstructed from mixed-spectral projection data via the FDK algorithm. $\eta$ and $\delta$ are weighting factors to balance the tradeoff between conventional and prior-based TV regularization terms.

**2.4 Implementation details**

All the experiments were implemented on an NVIDIA A100 GPU with a memory of 80 GB. The optimization of Eq. (5) was solved using an Adaptive Moment Estimation algorithm built in the PyTorch framework [41]. The step size was set to 0.001 in the physical phantom experiments and 0.1 in the digital phantom experiments. The first-order decay rate was 0.9, and the second-order decay rate was 0.999. The weighting factor of the SSIM term ($\lambda$) was set to 0.1 in all the experiments. In the implementation of the DTV method, the directions $i, j$ were selected based on the limited-angle artifacts in the FDK results and were indicated in the figures. $\eta_i$ was set to 0.5 and $\eta_j$ was set to 0.1 in all the experiments. In the implementation of the PICCS method, $\eta$ was set to 0.25 and $\delta$ was set to 0.75 in all the experiments. The reconstruction took 10 seconds in the physical phantom studies and 260 seconds in the digital phantom studies.

# 3. Results

## 3.1. Study of physical phantoms

Figure 1 summarizes reconstructed DE-CBCT images of the Catphan©700 phantom. The reference images were reconstructed from 360° projection data using the FDK algorithm. Obvious limited-angle artifacts appear in the results of FDK and DTV. Limited-angle artifacts are reduced in the images reconstructed using PICCS, but there are some residual artifacts along specific directions. The proposed method achieves efficient limited-angle artifact reduction, and no visible distortion is observed in the results.

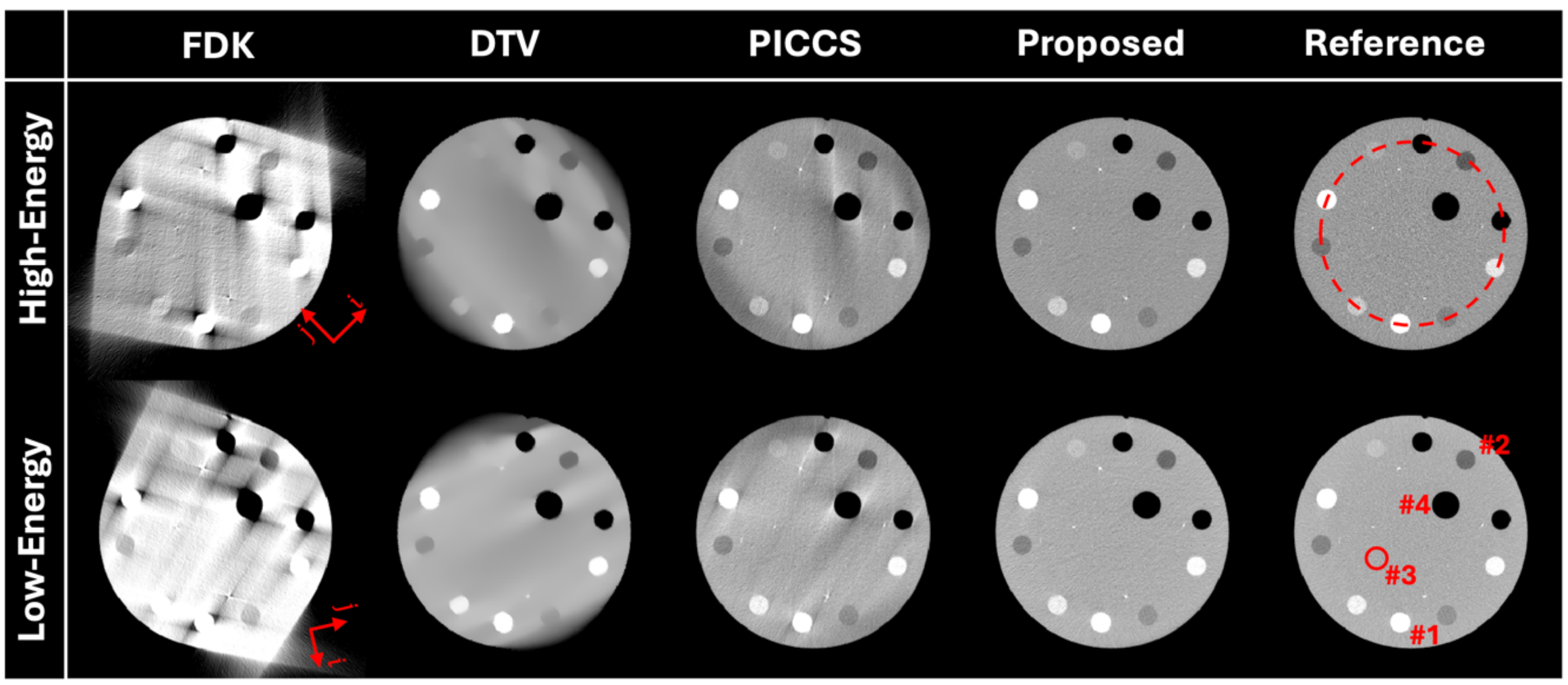

**Figure 1.** Reconstructed limited-angle DE-CBCT images of the Catphan©700 phantom using different methods. Display window: [-500 500] HU. The arrows in the FDK results indicate the directions of DTV regularization. The dashed circle and four ROIs in the reference images are used to measure line profile and HU accuracy.

The 1D profiles along the dashed circle indicated in Figure 1 are plotted in Figure 2, where the results of the proposed method show the best coherence to the ground truth.

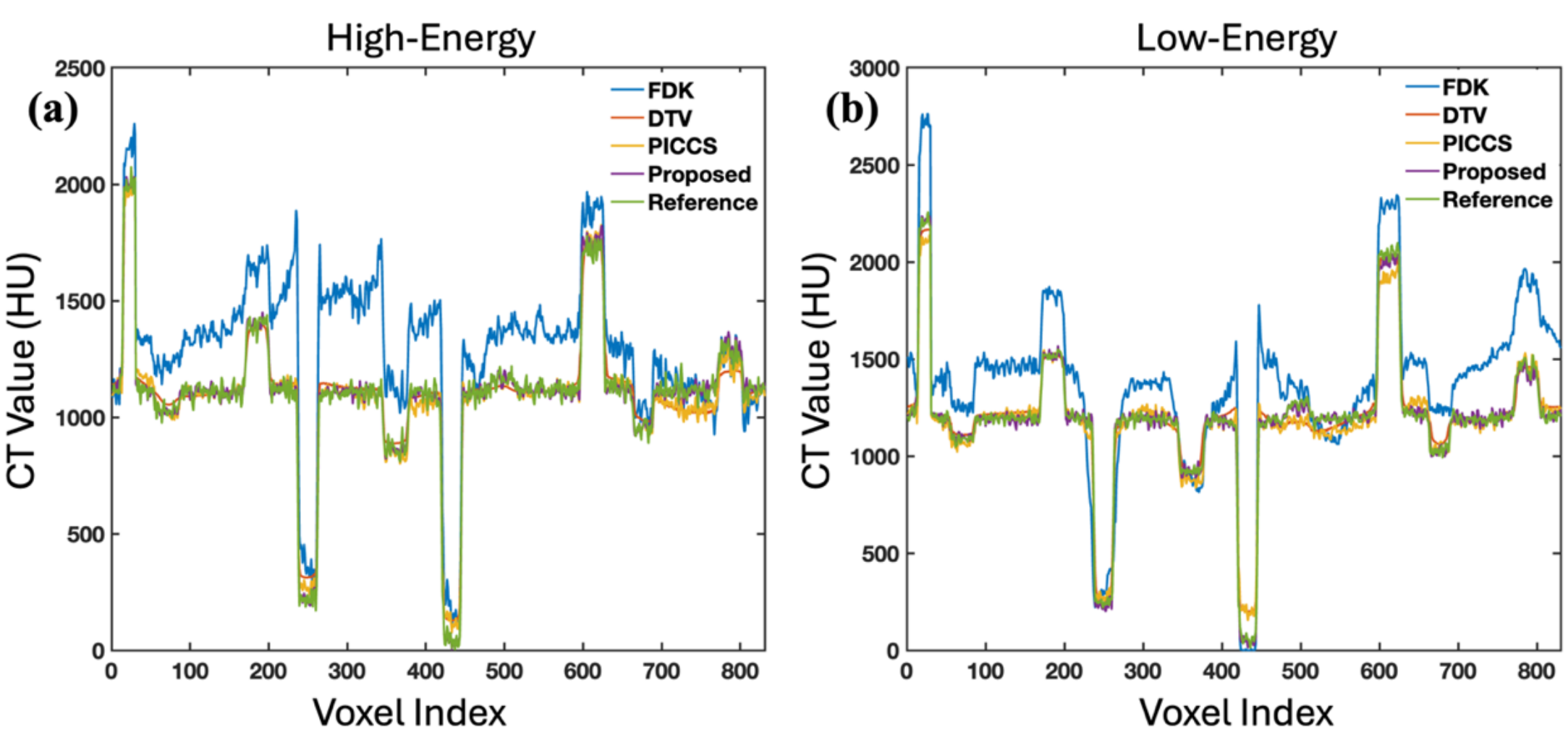

**Figure 2.** Line profiles along the dashed circle in Figure 1.

Four uniform ROIs in the Catphan©700 phantom were selected to measure HU accuracy, as indicated in Figure 1. The results are summarized in Table II, showing that the proposed method can achieve accurate image reconstruction in limited-angle DE-CBCT.

**Table II**. Mean and standard deviation of CT numbers (HU) measured within different ROIs which are indicated in Figures 1 and 3.

| | | Catphan©700 | | | | Head Phantom | |
|---|---|---|---|---|---|---|---|
| | | #1 | #2 | #3 | #4 | #1 | #2 |
| High-Energy | Proposed | 1001±15 | -137±12 | 101±13 | -959±11 | -15±11 | 212±9 |
| | Reference | 1003±65 | -136±58 | 104±70 | -967±65 | -14±67 | 209±71 |
| Low-Energy | Proposed | 1206±16 | -80±9 | 172±11 | -953±10 | 45±8 | 302±9 |
| | Reference | 1207±46 | -75±39 | 175±44 | -947±36 | 41±52 | 301±53 |

Figure 3 summarizes the results of the anthropomorphic head phantom. Consistent with previous results, the proposed method achieves the best performance of limited-angle artifact reduction, while obvious artifacts are observed in the results of other methods.

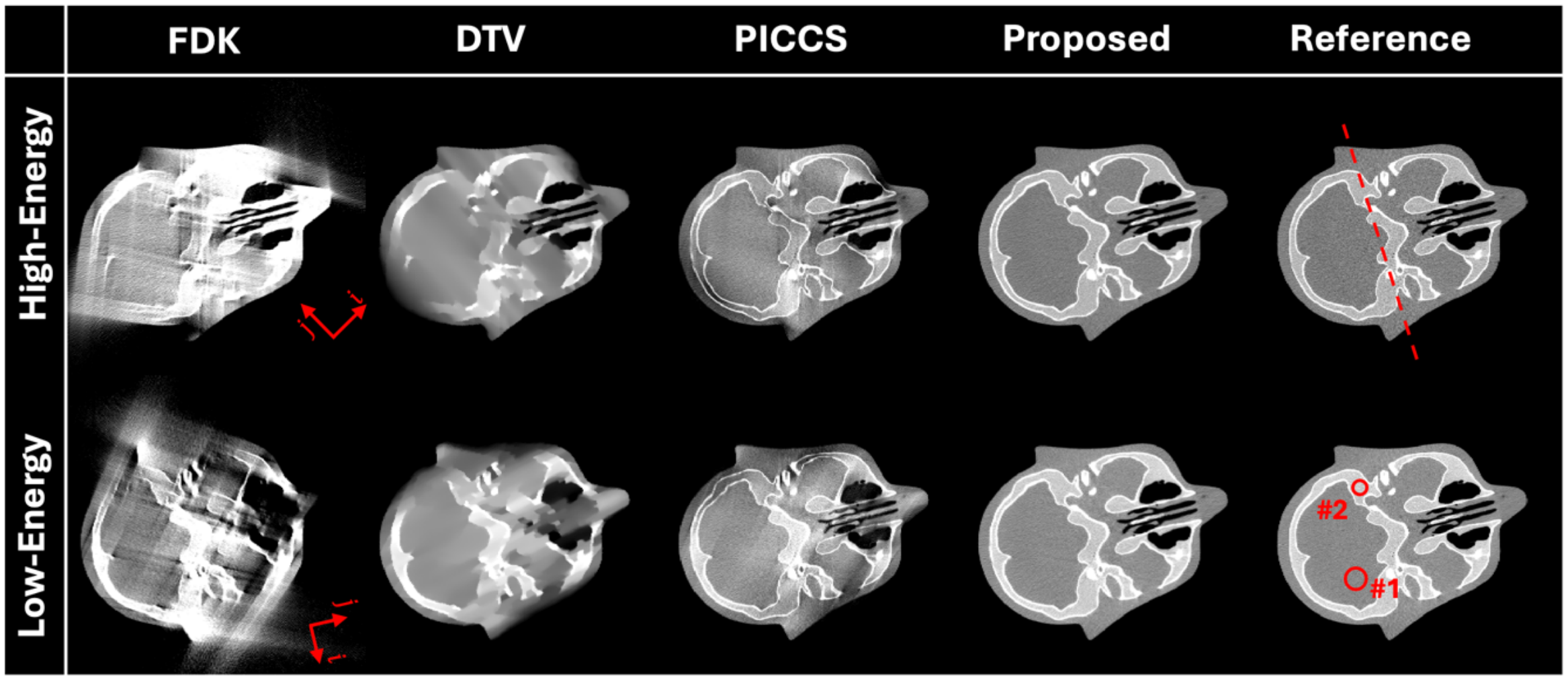

**Figure 3.** Reconstructed limited-angle DE-CBCT images of the Catphan©700 phantom using different methods. Display window: [-500 500] HU. The arrows in the FDK results indicate the directions of DTV regularization. The dashed line and two ROIs in the reference images are used to measure line profile and HU accuracy.

Figure 4 plots the 1D profiles along the dashed line indicated in Figure 3, where the proposed method shows the best coherence to the ground truth.

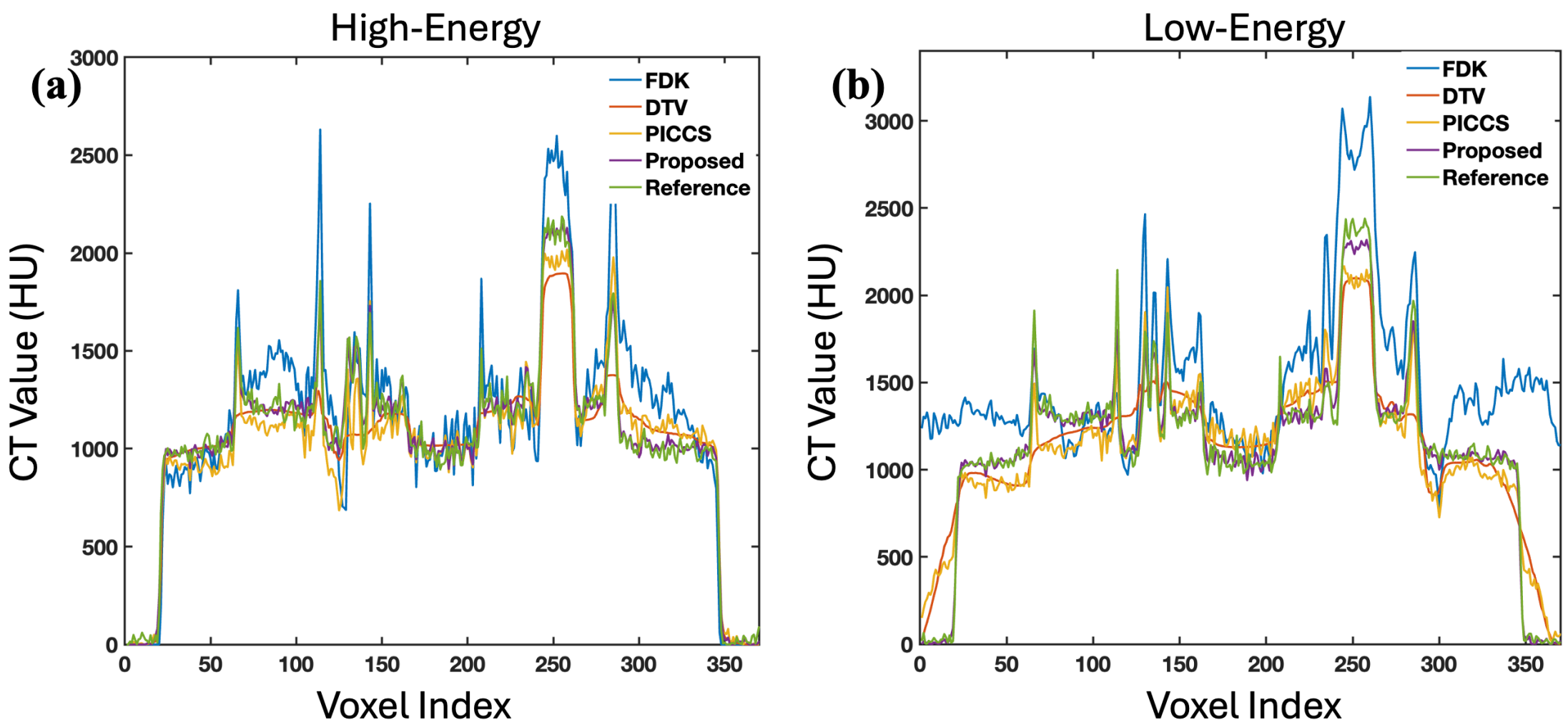

**Figure 4.** Line profiles along the dashed line in Figure 3.

The ROIs for HU accuracy measurement are indicated in the reference low-energy images in Figure 3, and the results are listed in Table II, showing great accuracy of reconstructed HU values using the proposed method.

Table III lists other quantitative metrics, including PSNR and SSIM, used in the study of physical phantom studies. The results of the proposed method show superior quantitative image quality over other results.

**Table III**. Quantitative metrics in the studies of physical phantom studies. The values are calculated between the reconstructed and reference images.

| | | | FDK | DTV | PICCS | Proposed |
|---|---|---|---|---|---|---|
| Catphan©700 | PSNR (dB) | High-Energy | 39.34 | 53.29 | 57.56 | 60.46 |
| | | Low-Energy | 39.20 | 53.51 | 57.29 | 62.30 |
| | SSIM | High-Energy | 0.71 | 0.98 | 0.99 | 1.00 |
| | | Low-Energy | 0.72 | 0.98 | 0.99 | 1.00 |
| Head Phantom | PSNR (dB) | High-Energy | 39.57 | 50.81 | 52.83 | 56.60 |
| | | Low-Energy | 39.48 | 48.97 | 50.77 | 56.31 |
| | SSIM | High-Energy | 0.71 | 0.98 | 0.98 | 1.00 |
| | | Low-Energy | 0.71 | 0.97 | 0.97 | 1.00 |

### 3.2. Study of digital phantoms

Figures 5-7 summarize reconstructed DE-CBCT images in different views of the three digital phantoms. The results using FDK and DTV show severe limited-angle artifacts, while residual artifacts are observed in the PICCS' results. The proposed method achieves efficient limited-angle artifact reduction, and no visible distortion is observed in the results.

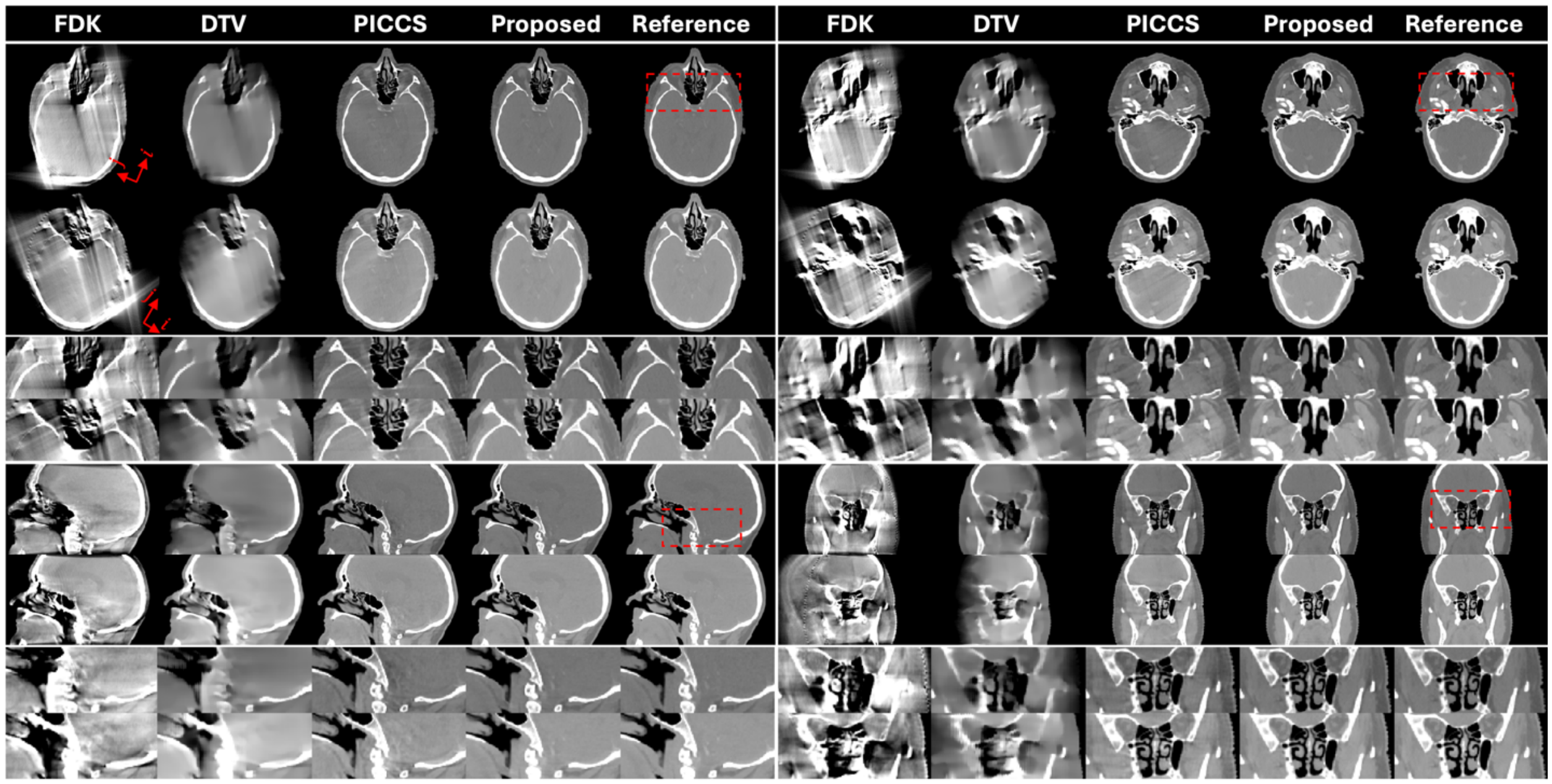

**Figure 5.** Reconstructed limited-angle DE-CBCT images of the digital phantom #1 using different methods. Display window: [-500 500] HU. The arrows in the FDK results indicate the directions of DTV regularization. Zoom-in images of the dashed boxes are provided to better demonstrate the limited-angle artifacts.

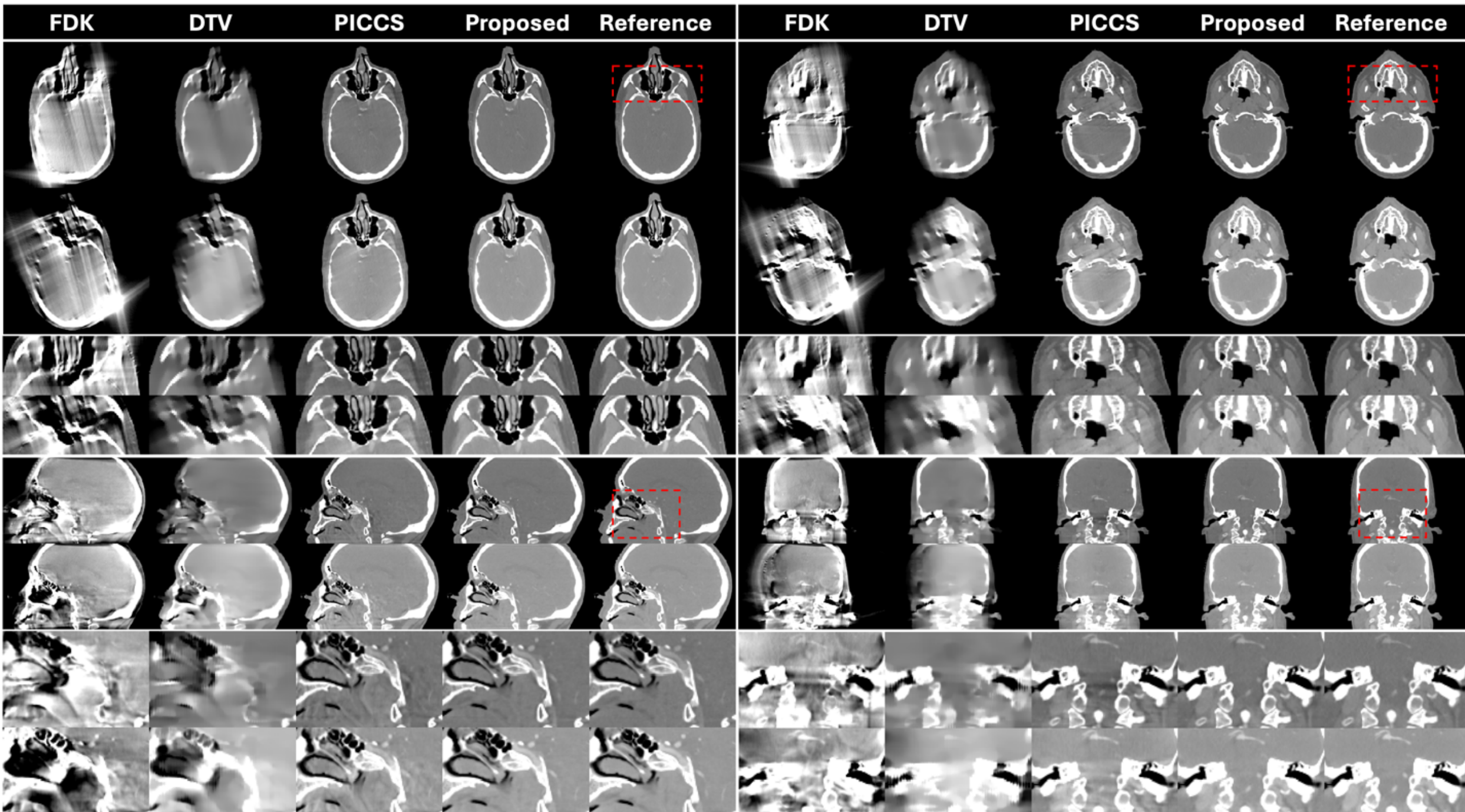

**Figure 6.** Reconstructed limited-angle DE-CBCT images of the digital phantom #2 using different methods. Display window: [-500 500] HU. The directions of DTV regularization are identical to the digital phantom #1. Zoom-in images of the dashed boxes are provided to better demonstrate the limited-angle artifacts.

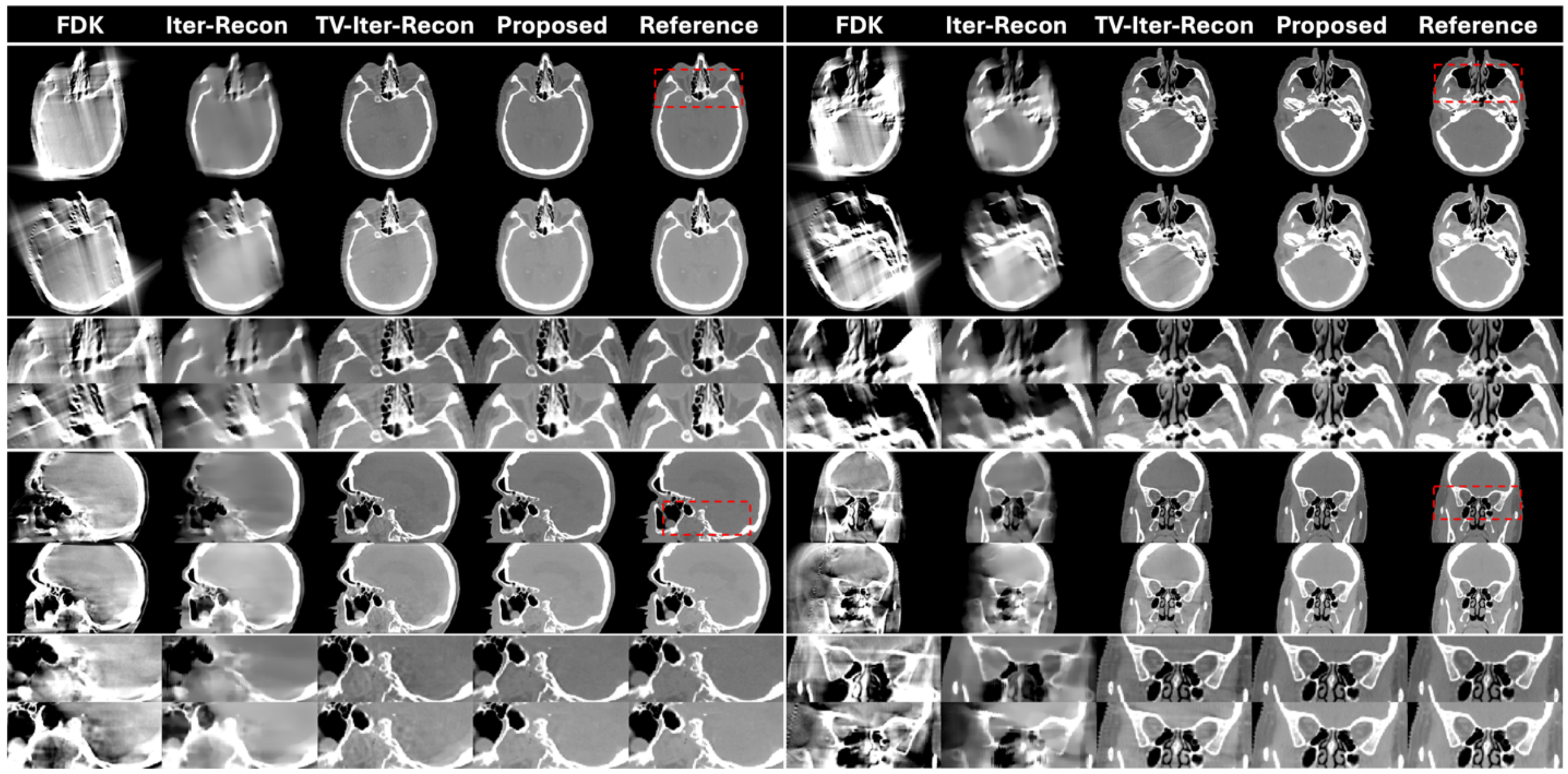

**Figure 7.** Reconstructed limited-angle DE-CBCT images of the digital phantom #3 using different methods. Display window: [-500 500] HU. The directions of DTV regularization are identical to the digital phantom #1. Zoom-in images of the dashed boxes are provided to better demonstrate the limited-angle artifacts.

Table IV summarizes the quantitative analysis results of the digital phantom study. The MAE is computed within the body mask. All the metrics are calculated for each slice, and the means and standard deviations are computed from all the 450 values.

**Table IV**. Quantitative analysis of the digital phantom study. The results are calculated between the reconstructed and reference images.

| | | FDK | DTV | PICCS | Proposed |
|---|---|---|---|---|---|
| MAE (HU) | High-Energy | 308.51±42.2 | 113.01±26.01 | 29.49±7.15 | 14.42±2.37 |
| | Low-Energy | 289.69±71.2 | 193.17±40.46 | 45.26±10.45 | 19.94±3.16 |
| PSNR (dB) | High-Energy | 40.20±0.74 | 51.64±1.30 | 65.04±1.14 | 69.88±0.96 |
| | Low-Energy | 39.15±0.72 | 47.73±1.54 | 60.95±1.14 | 66.93±0.76 |
| SSIM | High-Energy | 0.74±0.02 | 0.98±0.00 | 1.00±0.00 | 1.00±0.00 |
| | Low-Energy | 0.72±0.02 | 0.97±0.01 | 0.99±0.00 | 1.00±0.00 |

## 4. Discussion

This work presents a practical solution to image reconstruction in limited-angle DE-CBCT. By enforcing the inter-spectral structural similarity constraint between DE-CBCT images during the optimization-based image reconstruction, limited-angle artifacts are efficiently reduced in the reconstructed images.

Optimization-based image reconstruction algorithms, such as PICCS and DTV, have been introduced to relax the data acquisition requirement of DECT for a long time and have achieved great success. The main idea of PICCS-based DECT methods is to obtain a prior image with accurate anatomical structures and enforce the constraint in similarity between the target image and the prior image during the subsequent iterative reconstruction. In Ref [22], a prior spectrally average CT is reconstructed from the slow kVp-switching projection data and used for target DECT reconstruction from sparse-view data. In Ref [42], a prior high-energy CT is reconstructed from full-sampled data and used for similarity-based regularization in the target low-energy CT reconstruction from sparse-view data. However, PICCS may not be the optimal strategy for image reconstruction with the data acquisition scheme in limited-angle DE-CBCT. In a series of works, DTV has been integrated into the material decomposition framework to achieve accurate DECT reconstruction using convex or nonconvex optimization [31-34]. In these works, the correlation between DE projection data is fully exploited via the material decomposition model, and the DTV regularization is enforced on the virtual monoenergetic or material-specific images. However, the implementation of such strategies may be challenging in limited-angle DE-CBCT due to the measurement of X-ray spectra and the intensive computation required. In short, both PICCS and DTV are existing arts that have significant impacts in this field. The proposed method is a fundamentally different strategy designed for efficient image reconstruction in limited-angle DE-CBCT, which may bring added value to this field.

A requirement for the proposed method is that the total angular coverage of limited-angle DE-CBCT must meet the Tuy's data sufficiency condition [1], because the success of the inter-spectral structural similarity regularization in this work relies on the acquisition of complete anatomical structures. In other words, the total coverage of high- and low-energy projection views cannot be less than a short scan in the proposed work. However, a shorter scan is required for some clinical scenarios such as motion management, lower radiation dose, and avoidance of possible collisions between the gantry and the subject to be imaged. This would be the focus of our follow-up studies to extend the proposed method to more challenging data acquisition schemes.

In principle, the deep learning-based image prior can provide more information than the mathematically explicit image prior, e.g., the proposed inter-spectral similarity regularization. However, the major bottleneck of deep learning-based methods is the data pair requirement for supervised model training. Therefore, another focus of our future studies is to develop unsupervised alternatives to the existing models, which would pave the way for clinical applications of the deep learning-based limited-angle DE-CBCT. In our follow-up studies, we will investigate how to incorporate the proposed optimization-based image reconstruction into the deep learning methods for unsupervised enforcement of projection data consistency.

## 5. Conclusions

In this work, we propose an optimization-based image reconstruction regularized with inter-spectral structural similarity for limited-angle DE-CBCT. The proposed method can efficiently reduce limited-angle artifacts during the image reconstruction, enabling quantitative DE-CBCT with comparable data acquisition time and radiation dose to that of a single-energy scan on current onboard scanners without hardware modification. This work is of great clinical significance and can boost the clinical application of DE-CBCT in image-guided radiation therapy and surgical interventions.

## Acknowledgment

This research is supported in part by the National Institutes of Health under Award Number R01CA272991, R01EB032680, R01DE033512, P30CA008748 and U54CA274513.